\documentclass[sigconf]{acmart}

\AtBeginDocument{%
  \providecommand\BibTeX{{%
    \normalfont B\kern-0.5em{\scshape i\kern-0.25em b}\kern-0.8em\TeX}}}

\copyrightyear{2020} 
\acmYear{2020} 
\setcopyright{acmcopyright}
\acmConference[ICSSP '20]{International Conference on Software and System Processes}{October 10--11, 2020}{Seoul, Republic of Korea}
\acmBooktitle{International Conference on Software and System Processes (ICSSP '20), October 10--11, 2020, Seoul, Republic of Korea}
\acmPrice{15.00}
\acmDOI{10.1145/3379177.3388897}
\acmISBN{978-1-4503-7512-2/20/05}

\usepackage{todonotes}
\usepackage{amsmath,amssymb,amsfonts}
\usepackage{algorithmic}
\usepackage{graphicx}
\usepackage{textcomp}
\usepackage{xcolor}
\usepackage{url}
\usepackage{booktabs}
\usepackage{xspace}

\begin{document}

\title{Charting Coordination Needs in Large-Scale Agile Organisations with Boundary Objects and Methodological Islands}

\author{Rashidah Kasauli}
\orcid{}
\affiliation{
\institution{Chalmers $\mid$ University of Gothenburg}
\city{Gothenburg} 
\country{Sweden}\\
\institution{Makerere University, Uganda}}
\email{rashida@chalmers.se}

\author{Rebekka Wohlrab}
\affiliation{
\institution{Chalmers $\mid$ University of Gothenburg} 
\institution{Systemite AB}
\city{Gothenburg} 
\country{Sweden}}
\email{wohlrab@chalmers.se}

\author{Eric Knauss}
\affiliation{
\institution{Chalmers $\mid$ University of Gothenburg}
\city{Gothenburg}
\country{Sweden}}
\email{eric.knauss@cse.gu.se}

\author{Jan-Philipp Stegh\"ofer}
\affiliation{
\institution{Chalmers $\mid$ University of Gothenburg}
\city{Gothenburg}
\country{Sweden}}
\email{jan-philipp.steghofer@cse.gu.se}

\author{Jennifer Horkoff}
\affiliation{
\institution{Chalmers $\mid$ University of Gothenburg}
\city{Gothenburg}
\country{Sweden}}
\email{jenho@chalmers.se}

\author{Salome Maro}
\affiliation{
\institution{Chalmers $\mid$ University of Gothenburg}
\city{Gothenburg}
\country{Sweden}}
\email{salome.maro@cse.gu.se}

\renewcommand{\shortauthors}{Kasauli, Wohlrab, Knauss, Stegh\"ofer, Horkoff and Maro}

\begin{abstract}
Large-scale system development companies are increasingly adopting agile methods.
While this adoption may improve lead-times, such companies need to balance two trade-offs: (i) the need to have a uniform, consistent development method on system level with the need {for} specialised methods for teams in different disciplines (e.g., hardware, software, mechanics, sales, support); (ii)  the need for comprehensive documentation on system level with the need to have lightweight documentation enabling iterative and agile work.
With specialised methods for teams, isolated teams work within larger ecosystems of plan-driven culture, i.e., teams become agile ``islands''.
At the boundaries, these teams share knowledge which needs to be managed well for a correct system to be developed.
While it is useful to support diverse and specialised methods, it is important to understand which islands are repeatedly encountered, the reasons or factors triggering their existence, and how best to handle coordination between them. 
Based on a multiple case study, this work presents a catalogue of islands and the boundary objects between them.
We believe this work will be beneficial to practitioners aiming to understand their ecosystems and researchers addressing communication and coordination challenges in large-scale development. 
\end{abstract}



\maketitle

\section{Introduction}
\label{sec:introduction}
Large-scale systems engineering companies are typically made of many teams that work together, commonly with plan-driven methods, to contribute to one product. 
With the recent digitisation trends, many such companies have adopted agile methods to help them address the ever-changing market needs and the growing competition \cite{pernstal2012,van2013agile}. 
Given the challenges of introducing agility to large system companies \cite{lindvall2004agile}, most adoptions start with the software development teams at the lower level in the companies \cite{west2011water}. 
These teams in turn tailor the agile methods to their contexts \cite{Campanelli2015}, thus creating companies that have pockets of agile teams within a larger ecosystem of plan-driven culture, also identified as ``\textit{agile islands in a waterfall}'' \cite{kasauli2017requirements}.  

As the combination of agile with traditional plan-driven development methods become reality \cite{theocharis2015water}, knowledge management and coordination challenges {arise} \cite{van2013agile}. 
Inter-team coordination and knowledge sharing are key items on the research agenda on large-scale agile development~\cite{Dingsoyr2018}.
Teams using varying methods and practices need to communicate to deliver the correct product.
Artefacts shared at their boundaries, referred to as boundary objects, 
offer potential solutions to {these knowledge and coordination} challenge{s}. 
Boundary objects have been proposed to help manage coordination between agile teams~\cite{Wohlrab2019JSME}.
They can create a common understanding across sites, without compromising each team's identity{,} and are therefore useful when establishing coordination mechanisms across team boundaries.

To this effect, and as a first step towards alleviating the coordination challenge, this study set out to understand and document the agile islands together with the boundary objects that are constantly encountered {in large-scale systems development}.
Through an exploratory study with four large-scale {system engineering} companies, based on a focus group and two workshops, we explore the following research questions:

\newcommand{\researchquestionone}{Which agile islands are repeatedly encountered in large-scale agile contexts?\xspace}

\newcommand{\researchquestiontwo}{Which boundary objects are repeatedly encountered in large-scale agile contexts?\xspace}

 \emph{\textbf{RQ.1:} \researchquestionone}
 To understand {how best to address the coordination issue},
 we uncover the different islands that are encountered in the companies.
 
 \emph{\textbf{RQ.2:} \researchquestiontwo}
  To understand {how best they can coordinate and manage},
  we document the boundary objects that are shared between islands. 
 
{Interestingly, we find that not all islands reported are indeed \emph{agile islands} within a non-agile context. 
Significant distance can also occur between two agile teams and even the distance between two non-agile teams can have an impact on large-scale agile system development.}
{In addition, we find concrete methodological islands, but also more abstract forces (or: \emph{drivers}) that can contribute to the emergence of islands.}
 
In answering our research questions, we provide a catalogue of {methodological} islands that {are frequently relevant when introducing agility} at scale, {as well as} the boundary objects between them. 
We believe that this study will benefit both researchers and practitioners who want to gain insights into inter-team coordination in large-scale development.  

This paper is organised as follows: Section \ref{sec:background} presents the background to our study. In Section \ref{sec:method}, we describe the methods we used to answer our questions and gives the details of the workshops and focus group. Section \ref{sec:islands} presents our findings to RQ.1 and Section \ref{sec:boundary-objects} describes the findings related to RQ.2. We discuss our findings and conclude the study in Section \ref{sec:discussion}.

\section{Background} \label{sec:background}
Agile methods with the promise of continuous delivery of quality software have changed the way software is developed since the launch of the agile manifesto in 2001~\cite{Beck2001}. 
Originally meant for small co-located teams, agile methods are being adopted in large-scale systems development organisations  \cite{paasivaara2018large}. 
Existing studies on agile adoption in large-scale systems show that companies successfully adopt these methods \cite{salo2008agile,lindvall2004agile} even though challenges {remain}, {especially those related to coordination and mixed processes between different teams} \cite{scheerer2014coordination,Dingsoyr2018}.
This section gives the background of agile islands and boundary objects presented in this study. 

\subsection{Agile Islands}
Many large-scale {system engineering} companies have not fully adopted agile methods since they are not fully applicable in their domains.
Empirical researchers have recommended tailoring agile methods to the contexts of the specific organisation \cite{lindvall2004agile}. 
This means that organisations have to carefully choose practices that complement their values, culture, and norms \cite{qumer2008framework}. 
Research on agile methods tailoring has covered diverse areas including methods used to tailor \cite{Campanelli2015} and also the rationales and implications of tailoring \cite{Kalus2013}.
Still, {to satisfy the need to complement agile methods with traditional methods,} many companies are using hybrid methods in their development process \cite{theocharis2015water}.

Hybrid methods typically combine agile and plan-driven practices in software development \cite{kuhrmann2017hybrid}. 
Existing research on this topic has confirmed that this is the trend in many organisations today \cite{theocharis2015water, Kuhrmann2018}. 
Studies have explored the challenges faced in such environments \cite{van2013agile,kusters2017agile} and others have gone a step ahead to propose solutions \cite{kuusinen2016strategies}. 
Tell et al. \cite{Tell2019} have studied how different practices are combined to devise hybrid processes in an attempt to understand how to systematically construct synergies.

It should be noted that {in large-scale organisations in practice,} combinations start with the software development teams using agile methods while the rest of the organisation works with traditional methods \cite{west2011water}. 
This leaves teams as ``agile islands'' in a waterfall environment \cite{kasauli2017requirements}, also defined as pockets of agile within larger ecosystems with plan-driven culture.

{Vijayasarathy and Butler reason that the} choice of method used in the teams is associated with characteristics of the organisation, project, and team size \cite{Vijayasarathy2016}.
{This offers an explanation for the existence of agile islands, that differ from the surrounding organization, e.g., in terms of artefacts, iteration length, and delivery schedule.}
{Bjarnason et al. refer to such differences as} different forms of distances, for instance, geographical, organizational, or cognitive distance, distance related to artefacts (e.g., semantic distance), and distance related to activities (e.g., temporal distance)~\cite{Bjarnason2017}. 
{Such} distances makes it more difficult to coordinate between islands or between the non-agile {part of the} organisation and the agile islands.

\subsection{Boundary Objects}
Boundary objects are a sociological concept introduced by Star and Griesemer~\cite{Star1989} who studied how a shared understanding between interdisciplinary stakeholders can be established.
We refer to their definition of boundary objects as ``objects which are both plastic enough to adapt to local needs and the constraints of the several parties employing them, yet robust enough to maintain a common identity across sites''~\cite{Star1989}.
In the context of agile development, the ``parties [...] across sites'' are individuals with potentially different backgrounds and disciplines, typically forming organisational units (e.g., teams or departments).
These groups can flexibly interpret a boundary object and tailor it to their needs, while the group's identity and existing practices can be preserved~\cite{Abraham2013}.
While boundary objects originate from the field of sociology, they have also been studied in agile development contexts (e.g.,~\cite{Zaitsev2016,Blomkvist2015}).
In these contexts, boundary objects are artefacts (e.g., design specifications or user stories) that create a common understanding between agile teams~\cite{Blomkvist2015}.

In large-scale agile systems engineering, boundary objects are used between individuals from several sub-disciplines of systems engineering, who refer to concepts with different terminologies and are often located at different geographic locations~\cite{Wohlrab2019JSME}.
The groups using boundary objects need to be understood to enable knowledge management and inter-team coordination in an organisation.
Some organisational groups might be agile islands, working in different ways than others parts of an organisation.
\begin{table*}[!ht]
\centering
\caption{Descriptions of participating companies}
\label{tab:data-collection}
\begin{tabular}{@{}p{.12\textwidth}p{.855\textwidth}@{}}
\toprule
\multicolumn{2}{@{}l@{}}{\textbf{Focus Group}}\\
\midrule
Company A & Develops telecommunications products. Hardware development is largely decoupled from the software development. New hardware becomes available with a regular, but {low} frequency. Thus, the software development sets the pace of system development, which can be seen as continuous and agile, in that it embraces agile values as much as possible.\\
Company B & Develops mechanical products, both for consumer markets and for industrial development and manufacturing. Their system development is decomposed into several system elements. Software development is mostly confined to two of these elements, both of which are characterised by agile methods and practices such as Scrum and Continuous Integration. \\
Company C & Is an {automotive} OEM whose agile methods have been successfully applied to in-house development of software components. There is a desire to scale up these fast-paced approaches from developing software components to developing complete functions, thus including agile development of hardware and mechatronics.\\
Company D & Is a manufacturing company that develops high-tech products for the medical domain. Agile principles and practices are considered on all levels, yet must be carefully considered due to regulatory requirements and the very large scale of the development effort. The software development is to a good extent independent from hardware development cycles. \\
\midrule
        \multicolumn{2}{@{}l@{}}{\textbf{First Workshop (Company B)}}\\
        \midrule
       13 practitioners & Systems  engineers, project managers, test specialists, digital transformation managers,  and  business  developers.\\
       \midrule
        \multicolumn{2}{@{}l@{}}{\textbf{Second Workshop (Company A)}}\\
        \midrule
3 practitioners & Scrum master, architect, systems engineer\\
\bottomrule
\end{tabular}
\end{table*}

\section{Research Method} \label{sec:method}
Due to the exploratory nature of our research questions, we decided to conduct a multiple exploratory case study~\cite{Yin2008}.
We collected data in a staged process, using a {focus group}
with participants from several companies as a starting point and refining our data with in-depth workshops at two of the companies. 
Table \ref{tab:data-collection} presents short descriptions of the participating companies.
We report on our data collection, the way we analysed the information, as well as the threats to validity in the following.

\subsection{Focus Group}
We base the findings of this study on a focus group in which we discussed agile islands and the boundary objects that connect them with four practitioners, one from each of the {four participating companies}.  
Three of those practitioners had prepared presentations based on our instructions to help us explore the following issues:
(i) in terms of inventory, what knowledge is required on the island and what knowledge actually exists; (ii) in terms of infrastructure, what knowledge needs to be shared and what knowledge is actually shared; and (iii) in terms of process, how to facilitate learning, retrieving, capturing and applying knowledge.

The practitioners have high-level technical roles in the organisation (system architect, tooling and process specialist) and are thus accustomed to working with different islands within the organisation. They have also been working in these companies for several years and thus have a good grasp of the processes and the organisational structure. All four companies are large-scale systems engineering organisations with a predominantly agile software development approach and global distribution of developers.

The three presentations identified boundary objects commonly encountered in practice together with teams that use them. They provided a foundation for identifying common boundary objects and islands and were the starting point for discussions about the commonalities and differences between the organisations.

This information was collected by the researchers with extensive notes. One of the researchers also prepared an overview image of the boundary objects and the islands they connect and applied a rough clustering while the workshop was ongoing. This figure was continuously augmented with new insights during the presentations and updated during the discussion. At the conclusion of the focus group, the figure was presented and practitioners could comment on whether it represented their understanding.

\subsection{Individual Company Workshops}
As a follow-up of the focus group, we conducted individual workshops of approximately three hours each with two of the companies that participated in the initial data collection.
The workshops were conducted onsite at the companies and aimed to analyze concrete agile islands and boundary objects based on the inventory from the focus group. Two researchers were involved in each of the workshops and acted as moderators and facilitators.
We prepared a workshop instrument
(\url{https://rebrand.ly/workshop_BOMI}) to introduce the topic and guide through the workshops.

The workshops started with an introduction to boundary objects and agile islands and a statement of the goals.
The participants were then asked to individually brainstorm the boundary objects and agile islands they encounter in their work.
All input was recorded on post-it notes. Agile islands were then discussed and roughly organized on a wall.
Once a picture of the relevant islands emerged, participants then located boundary objects between the identified islands {creating a map}.
This map was then discussed and the practitioners reflected on the implications of the islands and how the boundary objects are currently being managed.

The {first {company} workshop} attracted a total of 13 practitioners who represented a number of roles: systems engineers, project managers, test specialists, digital transformation managers, and business developers. 
Representatives from the company first presented their current development process and the transformation that they are undergoing.
Afterwards, the two researchers introduced agile islands and boundary objects and defined the purpose of the workshop.
We then followed the procedure outlined above. 
However, after the collection of islands and boundary objects and the initial discussion of the map, we focused on a specific boundary object (``Product Requirement Specification'') that was deemed highly critical by the practitioners.
This provided additional insights into differences of governance processes within the organisation as well as the impact of organisational cultures in different parts of the company.

At the second workshop, three company participants attended, having the roles of Scrum master, architect, and systems engineer.
The workshop procedure outlined above was followed, starting with an introduction of the concepts and goals, and ending with reflections on the implications of the findings.
The focus lay on Interface Descriptions, Product Backlog, and Customer Service Requests, and collected relevant characteristics for them.

{In each workshop, }two researchers took detailed notes of what was being said as well as pictures of the post-it notes.
Directly after the meeting, reflections were written down to allow for easier analysis.

\subsection{Data Analysis}

\begin{figure}
\centering
\includegraphics[width=\columnwidth]{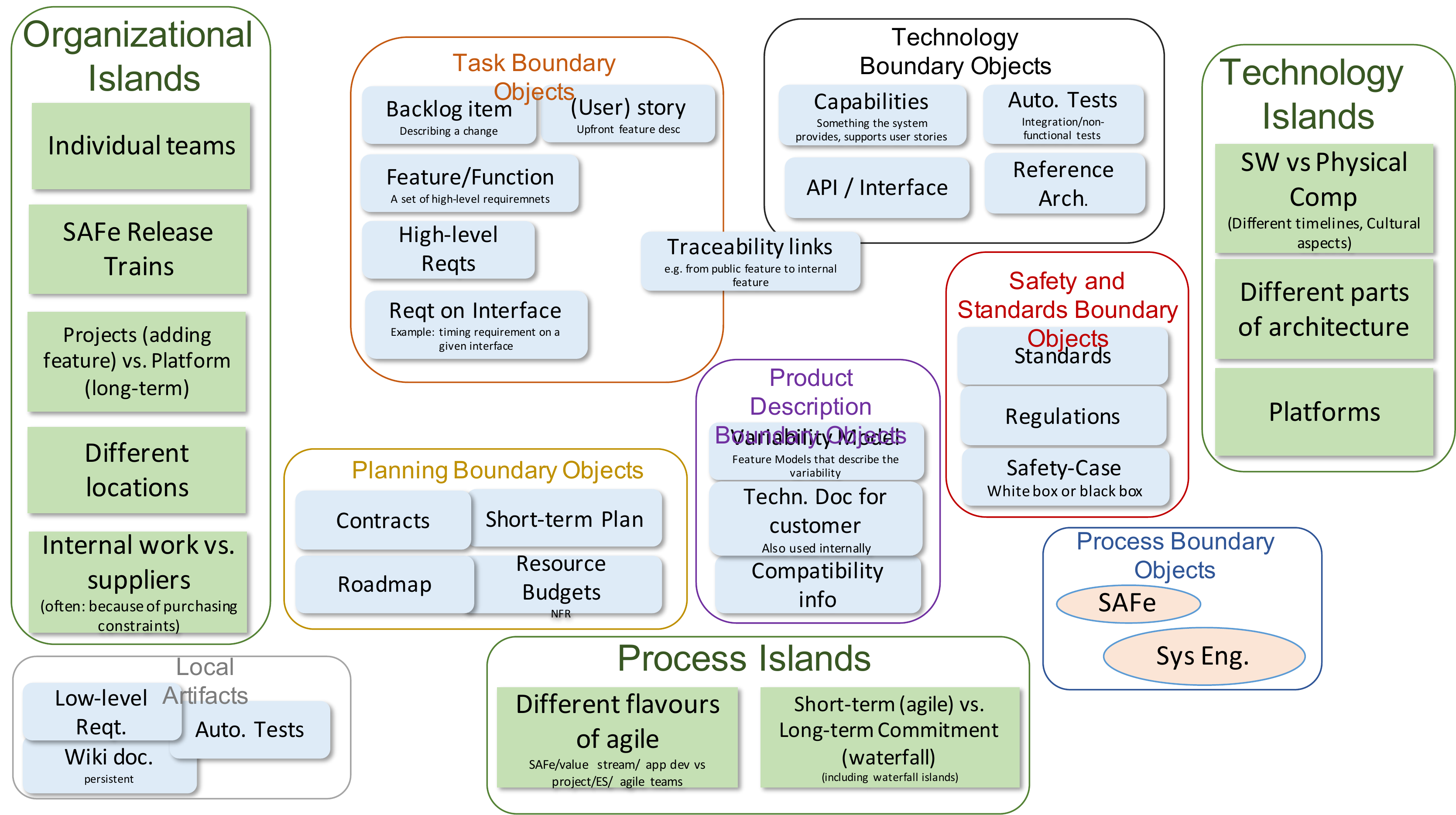}
\caption{Initial overview from focus group}
\label{fig:old-map}
\end{figure}

\begin{figure}
\centering
{\includegraphics[width=0.9\columnwidth, viewport=600 0 3000 2000, clip=true]{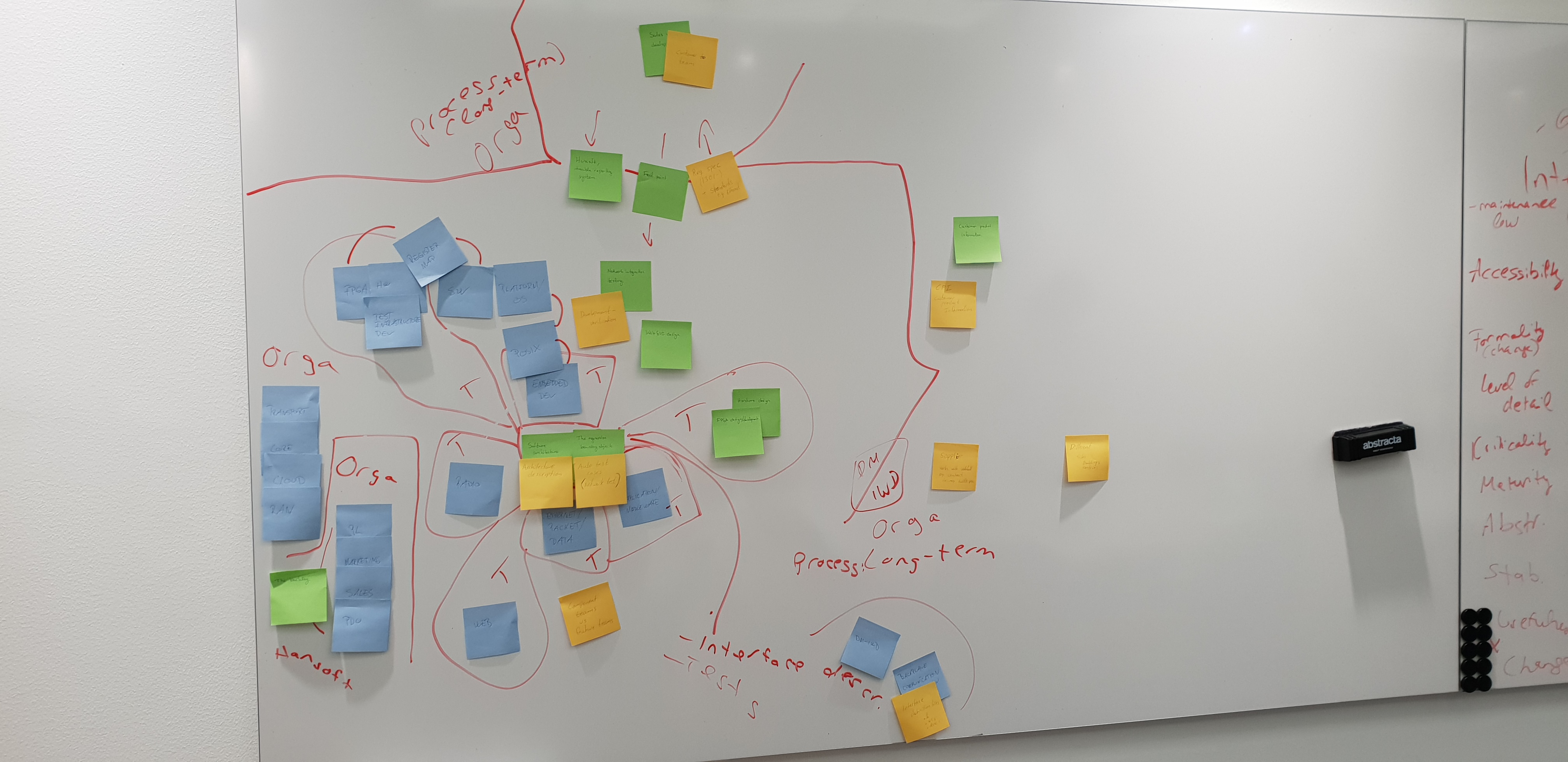}}
\caption{Brainstorming in second workshop (Company A)}
\label{fig:2ndWS}
\end{figure}

\begin{figure*}
\centering
\includegraphics[width=\textwidth]{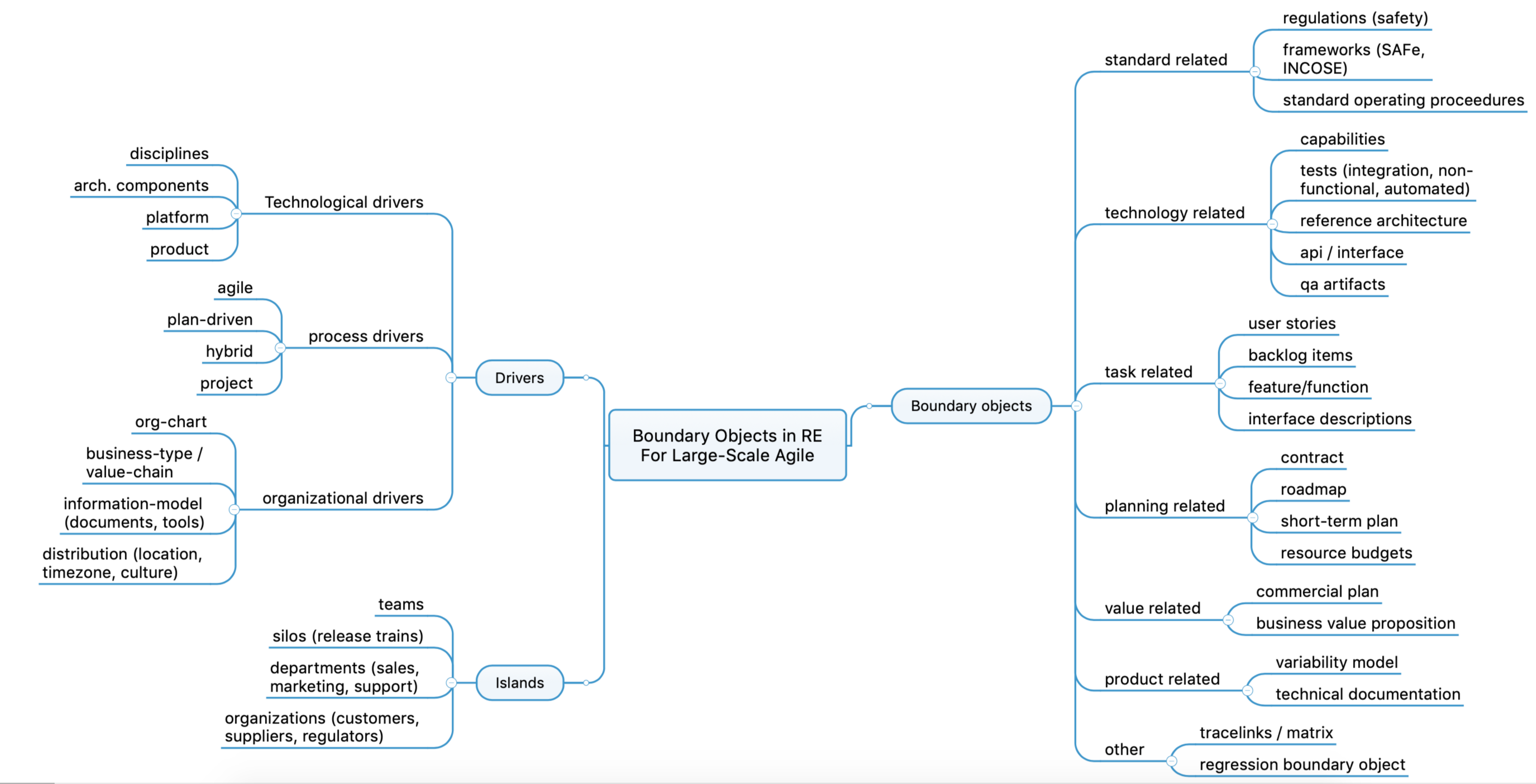}
\caption{First draft of results of Drivers, Islands and Boundary Objects}
\label{fig:new-map}
\vspace{-0.3cm}
\end{figure*}

All collected data was discussed between the researchers in groups. 
We used coding~\cite{liamputtong2009qualitative} to identify common themes in the agile islands and boundary objects we collected and structure the information in our transcripts and notes, as well as in the documents we collected from the practitioners. 
Discussions continued until an agreement about the codes was reached within the group of researchers. 
All findings were then member checked~\cite{de2017member} with the practitioners from whom the data was collected. 
The final results of these efforts provide the answers to the research questions outlined in Section~\ref{sec:introduction} and are presented in the following.
We collected the majority of our boundary objects and islands in the focus group.  The follow-up company workshops confirmed the existence of many of these islands and boundary objects, adding only a few new elements, increasing our confidence in our findings as per this set of companies.  
We demonstrate this process by including our initial overview after the focus group (Fig. \ref{fig:old-map}), a sample picture from the whiteboard after brainstorming with Company A (Fig. \ref{fig:2ndWS}), and a mindmap with the first draft of results reported in this paper (Fig. \ref{fig:new-map}).
In Figure~\ref{fig:new-map}, it can be seen that our initial findings were classified as boundary objects and islands, as well as ``technological drivers", ``process drivers", and ``organisational drivers". These findings were refined in several steps to arrive at the final results reported in this paper.

\subsection{Threats to Validity}

We addressed threats to \emph{internal validity} by including a number of practitioners in our workshops whenever possible and by allowing them to discuss their different perspectives on the data we collected. 
This increases our confidence that the data which forms the foundation of our study corresponds to the reality at the organisations that participated in it. 
The positive outcome of member checking our results further compounds this.

In terms of \emph{external validity}, we do not claim that our findings in terms of the concrete methodological islands and boundary objects we found are complete. 
By analysing data gathered from different companies with different characteristics, however, we believe that we have sketched out a framework that can be extended in the future and were able to identify relevant categories that are applicable in other contexts. 
It is our intention to extend the catalogue presented here and create a conceptual model of Boundary Objects and Methodological Islands (BOMI) with higher generalisability in the future.

To mitigate threats to \emph{construct validity}, we began each workshop with presentations explaining the concept of boundary objects and provided examples to help understanding.
The focus group targeted high-level experts from the respective companies. 
As reflected in their presentations, these experts understood the concepts well.
Also, questions were asked and clarifications made throughout the workshops.
{Thus,} all of our data collection tools focused on improving the understanding of the constructs under investigation, i.e., boundary objects and methodological islands. 
{Evaluation apprehension or experimenter expectancies are potential threats to construct validity. Peer debriefing helped us to critically reflect on these potential factors and the impact on our findings.}
To address \emph{reliability}, we combine a focus group with individual workshops at companies and combine the data collected in both to derive overall findings.

\section{Frequently Encountered Agile Islands (RQ.1)}
\label{sec:islands}

In this section, we present our findings with respect to 
{\emph{RQ.1 (\researchquestionone)}}. 
Overall, the discussion of agile islands resonated very well with our industry participants, both in cross-company workshops and in focus groups with individual companies. 
When analysing the collected data, we found a wide spectrum of relevant islands that we had to organize and categorize.
This led us to two observations: (i) not all islands that were mentioned are in fact \emph{agile islands}. Thus, there can be significant distance between two agile teams and even the distance between two non-agile teams can have an impact on large-scale agile system development. For this reason, we started to refer to the islands as \emph{methodological islands}. (ii) not all islands mentioned were on the same level of abstraction. While some (e.g., individual teams) are very concrete, others (e.g., ``software vs physical components'') are not very concrete agile islands, but can be seen as contextual factors that cause islands to emerge. We therefore started to refer to the latter as \emph{drivers of methodological islands.}

We first start to describe concrete methodological islands, before we also share the abstracted drivers. 

\subsection{Methodological Islands} 

The islands derived occur on different levels in the organisations. 
In Table \ref{tab:islands} we give an overview of levels and typical examples of islands.

\begin{table}[!hbt]
    \centering
    \small
    \caption{Examples of Methodologicical Islands on different levels.}
    \label{tab:islands}
    \begin{tabular}{@{}p{.2\columnwidth}p{.7\columnwidth}@{}}
        \toprule
        \multicolumn{2}{@{}l@{}}{\textbf{Groups of teams}}\\
        \midrule
        Groups of agile teams & Agile release trains or scrum-of-scum clusters may differ in methods.\tabularnewline
        Departments & Different methods and practices for SW development have emerged in different departments of large system companies (e.g., infotainment, powertrain in automotive) \tabularnewline
        Disciplines & Systems engineering needs to combine several disciplines, including hardware, mechanics, and software of different types, each with their own set of methods and practices. \tabularnewline 
        \midrule
        \multicolumn{2}{@{}l@{}}{\textbf{Individual teams}}\\
        \midrule
        Component teams & If teams are related to architectural components, they may favour different methods and practices.\tabularnewline
        Integration teams & Complex products may require dedicated support for continuous integration, provided by specialized testing and framework support teams. Their methods may differ significantly from other teams.\tabularnewline
        \midrule
        \multicolumn{2}{@{}l@{}}{\textbf{Organizations}}\\
        \midrule
        Suppliers & If an OEM aims for continuous integration, they may require suppliers to continuously deliver SW components. Naturally, methods and practices differ between customer and supplier and between suppliers.\tabularnewline
        Consultants & Systems engineering companies may rely on consultants to help developing software components. These may again bring a different set of methods and practices. \tabularnewline
        Regulators & Agile system development of regulated systems needs to take into account methods and practices of regulators. These may differ between domains and particular regulators.\tabularnewline
        \bottomrule
\end{tabular}
\end{table}

\subsubsection{Groups of teams}
Two of our participants' companies implemented the Scaled Agile Framework (SAFe). SAFe suggests the use of Agile Release Trains, i.e., of a team of agile teams that together develop and deliver a solution.
Value streams exist on the highest level of SAFe.
Within each value stream, there are multiple release trains.
In one of the participating companies, there are about 50 release trains with 5 to 12 teams in total.
Internally, these agile release trains require synchronisation and coordination, but externally, they can be perceived as a black box. 
These release trains develop different (sub-)systems that have interfaces with each other.
When several release trains depend on each other, their differences in methodology become an obstacle.
Officially, departments are not mentioned anymore in the SAFe-related documentation, but have traditionally existed in the companies.
Release trains are orthogonal organisations to the former departmental structures and can span several departments in the company.

Product development {typically} spans several departments in an organisation. 
These departments, for example, marketing, hardware development, embedded system development, come from different contexts and thus different {ways of working}. 
As it was not yet clear how hardware can work in an agile way {or if they even should}, the hardware teams for instance maintained {plan-driven methods} and yet they have to interface with software teams that are already adopting agile methods.
Hardware and software departments work using different timelines. 
It is also common for globally distributed companies to have departments spanning different locations that could spur different methods within the department due to the difference in cultures. 
Each department can have several teams with a common goal of contributing to a single component or feature of the product. 

\subsubsection{Teams (individual teams)}
Within an organisation, different teams can follow different agile approaches or even work in an agile way while the rest of the organisation follows a plan-driven approach.  
Teams in such large companies handle different parts of the architecture of the product. 
This means that each team works with different requirements and thus could use different approaches to get to the solution.
Participants mentioned, for instance, continuous Integration framework teams, integration testing teams, {Web GUI teams}, and software teams{. All of these teams may contribute to the same product, but since the nature of their tasks differs significantly, they often tailor development processes to their needs.}
{This leads to a set of methodological islands throughout the organization}.

\subsubsection{Organizations}
Companies work with suppliers, customers and regulators all of which come with different ways of working from that of the corresponding company. 
The suppliers provide some components while others are developed in-house. 
When teams within an organisation rely on external suppliers for components, the supplier is often working in a waterfall way. 
For instance, contracts between both companies often imply a plan-driven approach since purchasing is based on clearly defined functionality to be delivered at a certain point in time.
Regulators also rely on standards that do not explicitly specify the methods to use in development, but come with checkpoints that relate mostly to the plan-driven methods of working. 
This mismatch of the actual methods used versus the 'unknown' expectations becomes a hindrance in development. 

\subsection{Drivers of Methodological Islands}
\label{sec:drivers}
The methodological islands are triggered by certain factors that we derived upon analysis. 
We summarize these in Table \ref{tab:drivers} and describe them below. 
\begin{table}[!hbt]
    \centering
    \small
    \caption{Different types of drivers for methodological islands.}
    \label{tab:drivers}
    \begin{tabular}{@{}p{.3\columnwidth}p{.65\columnwidth}@{}}
        \toprule
\textbf{Business-related}
        &
        Economic function, Characteristics of market or value-chain, global distribution \tabularnewline
        \textbf{Process-related} & Mixture of development methods (SAFe, V-Model, Scrum, Kanban, LSD); focus on projects or products\tabularnewline
        \textbf{Technology-related} & Architectural decomposition, systems disciplines, platform and product-line strategy, time-scale of commitment \tabularnewline
        \bottomrule
        \end{tabular}
\end{table}

\subsubsection{Business-related drivers}
{Based on their history and business domain,} companies have {specialized} organisational chart{s} that describe the departments, e.g., for marketing, development, verification.
These departments handle different parts of the product that in most cases {imply} varying needs{ for development methods}. 
For instance, the sales department as opposed to development departments, have different needs and thus different ways of working. 
This difference, in turn sparks the need to adopt the agile practices to the context of the {specific} department, causing islands of {methodology, for example, manifesting in different choices with respect to forming cross-functional component or feature teams}. 
{Such business drivers can be the result of a particular culture in a market or value-chain.
How is the relationship between customers and suppliers characterised in terms of contracts, time-lines, trust, and interaction?
To what extent are customers willing or able to assess and give feedback on frequent deliveries?
Is it possible to take end-user opinions into account and to what extent do they matter?}
All these {aspects} contribute to the mix of methods and how the other stakeholders are going to work to get the product they need. 

Some of our participant companies are distributed over several countries and in some cases different areas in the same country. 
Developers of software or hardware  do not work in the same buildings and are separated by location, time zones and culture. 
This separation in the end creates teams that have defined different methods of doing the same thing. 

\subsubsection{Process-related drivers}
While teams exist in the organisation and have varying needs, the organisation in the end has to have one {backbone process} that defines the company. 
It is not uncommon that different teams within the same organisation use different flavours of agile methods. 
Apart from process customisation for each individual team, it is possible that some teams use a method such as SAFe while others employ Scrum, Kanban, XP, or a form of lean development. 
These differences introduce islands where roles, {artefacts}, and schedules are difficult to coordinate.

{A major driver of this category relates to whether a company mainly works based on projects, or whether significant work flows in the continuous development of a platform.}
While {projects} are adding features to a solution and are thus short-term, platforms are planned for the long-term. 
Platforms need to be more stable since other projects depend on them and changes in the platform can have a major impact on the depending projects.
{Thus, the particular setup of a company can create islands between different projects, or between customer projects and platform development.}

\subsubsection{Technology drivers}
Complex systems are often developed by different teams that are responsible for individual parts of the architecture. That means that these teams not only address different sets of requirements, but also apply different technologies in their solutions. 
Teams working on software and on physical components work according to different timelines and according to different cultures. 
Hardware development often assumes stable requirements and development of a full solution, instead of development of slices of functionality and rapid response to changes.

Many companies with complex product lines, e.g., in the automotive domain, produce platforms as the foundation of their products. 
Platforms are often generational, i.e., they are used for a certain period of time before they are replaced by the next generation. 
Each platform has a unique technical solution and is usually not compatible with previous ones. 
At the same time, different teams working on different platform generations also often use different generation of processes.

The time scale of commitment is another technology-related driver.
Agile methods usually imply short-term commitment in individual sprints. That means that requirements can change from sprint to sprint to react to a changing market situation or newly discovered opportunities. On the other hand, many methods require a longer-term commitment. Platforms, e.g., that are used by many other projects and thus need to be stable might be better served using a plan-driven approach and to constitute ``waterfall islands'' within the organisation.

\section{Boundary Objects in Large-Scale Agile (RQ.2)}
\label{sec:boundary-objects}
\begin{table*}[!ht]
    \centering
    \small
    \caption{Identified boundary objects and their categories.}
    \label{tab:boundary-objects}
    \begin{tabular}{@{}p{.2\textwidth}p{.77\textwidth}@{}}
        \toprule
        \multicolumn{2}{@{}l@{}}{\textbf{Task Boundary Objects}}\\
        \midrule
        Backlog item & Backlog items, e.g., from a product backlog, can be representations of high-level requirements and are used by individual islands to define their own, local backlog items for their product or team backlogs. \\
        (User) story & A user story is an upfront feature description focused on customer value. Backlog items can be formulated as user stories to clarify the value provided by delivering a piece of functionality. \\
        Feature, function description, or high-level requirements & A set of high-level requirements can be represented by a feature or function description. These requirements need to be further broken down to allow individual islands to work on them. \\
        Requirements on interfaces & Different parts of a software architecture are connected by interfaces. The requirements for these interfaces define contracts between teams. For instance, timing requirements on an interface need to be adhered to by all islands using this interface. \\
        \midrule
        \multicolumn{2}{@{}l@{}}{\textbf{Technology Boundary Objects}}\\
        \midrule
        Capabilities & A description of the capabilities provided by the system gives a high-level overview of the functionality. It allows individual teams to identify relevant reusable assets and required interfaces. \\
        Automated tests & Integration, acceptance, and non-functional tests can be shared between islands to avoid regressions, ensure customer value is jointly achieved, and to document the functionality provided in the system.\\
        API / Interface & The description of the interfaces between different parts of the solutions allow to modularise the development and different islands to reuse existing assets. \\
        Reference architecture & A high-level description of the architecture both allows different islands to identify where a feature should be located and ensures that new additions to the solution follow the common guidelines of the organisation. \\
        \midrule
        \multicolumn{2}{@{}l@{}}{\textbf{Regulation and Standards Boundary Objects}}\\
        \midrule
        Standards & Safety standards such as ISO~26262, DO-178B, or IEC~62304, prescribe development practices and artefacts.\\
        Regulations & Regulations take the role of standards and prescribe certain practices or artefacts (e.g., in the telecommunications domain).\\
        Safety assurance case & Safety standards prescribe the creation of white box or black box safety assurance cases that describe how a product addresses risks during its operation. These cases can be used by different islands to understand the risks involved in the system and to develop common strategies to avoid them or deal with them.\\
        \midrule
        \multicolumn{2}{@{}l@{}}{\textbf{Product Description Boundary Objects}}\\
        \midrule
        Variability model & The features of a product and the constraints between them (e.g., which ones are mutually exclusive or incompatible) can be used by different islands to understand the interaction between their solutions and the rest of the product line. \\
        Technical documentation for customer & Outwards-facing documentation can also be used internally to gain a common understanding of how different parts of a system are related. \\
        \midrule
        \multicolumn{2}{@{}l@{}}{\textbf{Process Boundary Objects}}\\
        \midrule
        SAFe documentation & The Scaled Agile Framework (SAFe) has found widespread adoption in large development organisations. It provides detailed documentation and support for its adoption. This documentation can, together with a description of how SAFe was adapted, act as a boundary object between islands using SAFe and other parts of the organisation not using SAFe. \\
        \midrule
        \multicolumn{2}{@{}l@{}}{\textbf{Planning Boundary Objects}}\\
        \midrule
        Contracts & The interactions between parts of an organisation and the suppliers are often defined by contracts. Contracts can also bind an island within an organisation to external constraints. In any case, the content of the contract will define the scope or the time and resources the island has at its disposal.\\
        Roadmaps & The long-term evolution of a product is often defined by one or several roadmaps. These boundary objects also link different products that co-evolve to each other. Therefore, they are used to coordinate between islands within an organisation. \\
        Short-term plans & The development of individual features or smaller parts of a product is often bound to a shorter-term plan that is connected to the overall, long-term plan. As such, the scope of a short-term plan is also limited to a smaller number of islands.\\
        Resource budgets & When developing systems in which software runs on dedicated hardware, individual islands need to work with a resource budget that determines how much computing power, memory, or bandwidth their specific functionality can consume. \\
        \midrule
        Trace links & Artefacts created during development need to be connected to each other using trace links. They clarify the relationship between artefacts and enable change impact analysis and collaboration between the islands that created the artefacts. \\ 
        \bottomrule
    \end{tabular}
\end{table*}

In this section, we answer \emph{RQ2: \researchquestiontwo}

Table~\ref{tab:boundary-objects} shows our findings for RQ2.
Each row represents a type of boundary object with a description of how it facilitates coordination between islands.
In total, 19 types of boundary objects were identified.
We categorized them in different themes: Task, technology, regulation and standards, product description, process, planning boundary objects, and trace links. 
We refer to the descriptions of the boundary objects in Table~\ref{tab:boundary-objects} and briefly summarize the categories in the following. 

\textit{Task boundary objects} relate to tasks in the development effort in which boundary objects facilitate the creation of a common understanding across team borders.
Concretely, these tasks are concerned with identifying development activities by creating a backlog and specifying requirements to define the functionality to be developed. 
{Typical examples are user stories and other backlog items as well as related comments stored in issue trackers.}

\textit{Technology boundary objects} are concerned with technological aspects of the (software) system to be developed, including a system's capabilities, tests, or architecture boundary objects.
These boundary objects are commonly used between different islands and mostly by technical stakeholders.

\textit{Regulation and standards boundary objects} are used to ensure that the company complies with regulations and standards.
In our case companies, it relates mostly to safety standards, as with the safety assurance case.
These regulations and standards are typically relevant across island borders and a common understanding of these concerns is required.

\textit{Product description boundary objects} relate to  the product as it will be sold to the customer.
While the respective documentation is mostly created for customers, it can also be leveraged internally, for instance, to create a shared understanding of variability concerns or other technical aspects.

\textit{Process boundary objects} are concerned with documentation regarding processes or frameworks.
In two of our case companies, SAFe is used and tailored to each company's needs. 
The created documentation can help to get a shared understanding of the processes and roles.

\textit{Planning boundary objects} relate to contracts, roadmaps, plans, or budgets that are created before development.
These boundary objects are commonly used between non-technical stakeholders like managers, but can also be relevant for development teams.

\textit{Trace links} are a special category, as they represent the relationships between artefacts.
Trace links typically have types that determine how other artefacts relate to each other.
They can also serve as boundary objects between different methodological islands{, capturing a mutual agreement about relatedness of other boundary objects.}

\section{Discussion and Conclusion} \label{sec:discussion}

\begin{figure*}[!ht]
    \centering
    \includegraphics[width=0.9\textwidth]{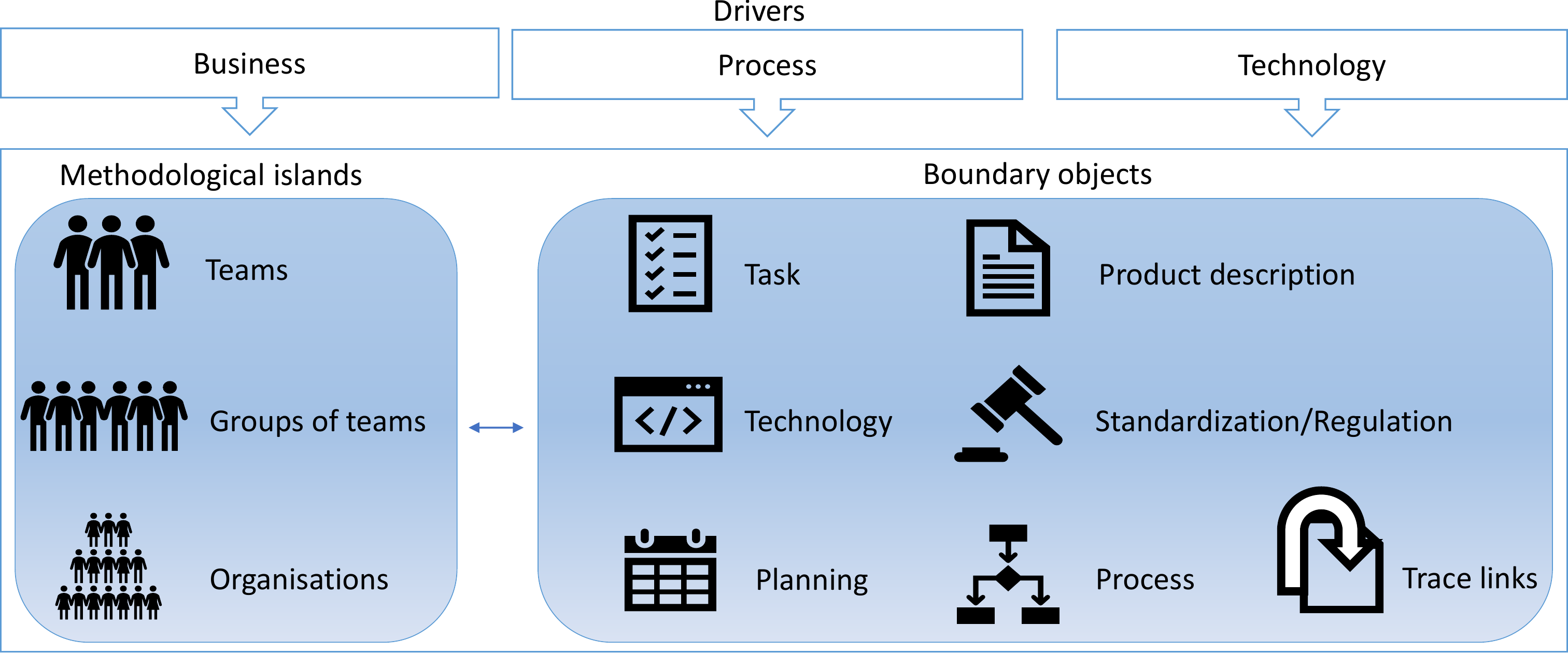}
    \caption{Summary of findings: Certain drivers introduce distance, which in turn frequently introduces methodological islands on different levels. Boundary objects of various types can be crucial to bridge between islands and support effective agile system development at scale.}
    
    \label{fig:summary}
\end{figure*}

In this paper, we presented methodological islands and boundary objects related to large-scale systems development collected from two workshops and a focus group with four large organisations.  
We present a summary of our findings in Figure \ref{fig:summary}.
As the first step towards addressing the coordination challenge in transforming organisations, we believe that this  study adds significant value both to research and to other organisations customising agile. We discuss our main findings and implications of our work in this section.

\subsection{Methodological Islands}
We discuss \emph{RQ.1 (\researchquestionone)} in this section.

Our findings show that {when embracing agile in large-scale system development, certain types of \emph{methodological islands} frequently appear on the level of individual teams, groups of teams, or full organisations.} 
Although not particular to large-scale, West et al.~\cite{west2011water} found that \emph{water-scrum-fall} is becoming a reality for most organisations, a claim confirmed by Theocharis et al.~\cite{theocharis2015water}. 
While terming them hybrid methods, Kuhrmann et al. \cite{kuhrmann2017hybrid} find that such hybrid approaches are not limited to traditional and agile development but also allow combinations of different agile methods since agile is also not implemented as is.
Tell et al.~\cite{Tell2019} go a step further and identify the agile methods and how they are combined in practice to form hybrid methods. 
Our findings on methodological islands confirm their findings as well as recognising that such combinations differ within the same organisation, causing methodological islands. 
Such islands are characterised by their relative distance in terms of methods and practices as well as culture and mindset.

In addition to the methodological islands, we found that certain drivers (business-, process-, and technology-related) can introduce such distance and lead to the formation of methodological islands.
This finding concurs, to some extent, with the finding by Vijayasarathy and Butler~\cite{Vijayasarathy2016} who found specific organisational, project and team characteristics had an effect on the choice of methodology. 
While we can confirm several of those characteristics, we come from the perspective of islands and classify the drivers as business-related, technology-related and process-related.  
{Team characteristics could	play a role but for the islands context, these are overtaken by e.g., the nature of (sub-)systems that different teams may be developing and thus we relate that driver to technology.}

\subsection{Boundary Objects}
In this section, we discuss the findings of \emph{RQ2: \researchquestiontwo}

{In order to successfully introduce agile methods and to deliver a full product or system, we found that effectively bridging between such islands is crucial. 
We believe that it is beneficial to think about artefacts that support such bridging as \emph{boundary objects} and provide in this paper an inventory of frequently encountered boundary objects.}
Many of the identified boundary objects have been confirmed by related studies.
In an analysis of boundary objects in distributed agile teams including developers and user-centered design specialists, Blomkvist et al.~identified the following boundary objects: (1) Personas, (2) Scenarios, (3) Effect maps, (4) Sketches, (5) Design Specifications, (6) Prototypes, (7) Evaluation summaries, and (8) User stories~\cite{Blomkvist2015}.

Our findings include a System Wiki boundary object, identified by company A.
Similarly, Yang et al.~\cite{Yang2008} name the use of a wiki as a boundary object for requirements engineering.
The accessibility and ability to simultaneously access and create information make wikis a suitable form for a boundary object.

In an analysis of boundary-spanning activities with a focus on requirements engineering practices for product families, {examples boundary objects included} traceability documentation, process models, vocabularies, user stories, product/process repositories, XP practices, feature models, the product backlog, the sprint backlog, and product prototypes~\cite{Jain2014}. 

In the area of requirements engineering, another study has focused on classification schemes as boundary objects, allowing stakeholders to categorise requirements in different ways (main users, functional vs.~non-functional, level of abstraction)~\cite{Hertzum2004}.
In fact, standardised forms and classification schemes have been examined in the context of boundary objects since their initial definition~\cite{Bowker1999,Star1989}. 
The regulation and standards boundary objects that we identified in this study relate to this category.
Moreover, process boundary objects potentially include classification schemes, for instance, by defining requirements information models that determine how stakeholders should work with requirements-related concerns and how they should be categorised~\cite{Wohlrab2020}.

Focusing on software development, project management documents and specifications have also been identified as boundary objects~\cite{Barrett2010}.  
{Thus our findings confirm many existing objects, and create a more integrated, industry-driven view of such objects in a large-scale agile context.}

\subsection{Implications for practitioners}

We found our inventory of methodological islands and related boundary objects useful when discussing potential process improvements with companies. 
Already the focus group and company workshops showed that this facilitates a useful mindset, where artefacts are discussed as a means to satisfy coordination needs between methodological islands. 
By making the islands explicit and by discussing their particular context, mindset, and preference with respect to methods and practices, we believe that such boundary objects can be established in a better way than if they would emerge in an unplanned way, e.g., by re-using non-agile artefacts.
Future work should investigate if this can be used constructively, when defining or improving processes, methods, and tools.

\subsection{Implications for research}
Similarly, we hope that charting the landscape of methodological islands and boundary objects in large-scale agile system development will create a useful model to scope and prioritize future research.  
Future research could refine the classifications in our inventory and provide a suitable {conceptual model or taxonomy}.  
{We are currently working with our participating companies to derive possible recommendations and best practices for boundary objects based on their properties.}
{In addition, a} quantitative study could {provide additional information on which boundary objects and methodological islands are most frequent}.

\begin{acks}
This work is partially funded by Software Center, Project \#27 ``RE for Large-Scale Agile System Development'' \url{www.software-center.se} and  the  Sida/BRIGHT  project  317  under  the Makerere-Sida bilateral research programme 2015-2020. 
We thank our industry partners for their enthusiasm and support.

\end{acks}

\bibliographystyle{ACM-Reference-Format}
\bibliography{boundary-objects}

\end{document}